\documentclass[]{raa}            

\usepackage{graphicx,times}             
\usepackage{natbib}
\usepackage{amssymb,amsmath}
\usepackage{longtable}
\bibpunct{(}{)}{;}{a}{}{,}

\usepackage[pagebackref=true]{hyperref}

\begin{document}

  \title{A statistical study of magnetic flux emergence in solar active regions prior to strongest flares}

   \volnopage{Vol.0 (20xx) No.0, 000--000}      
   \setcounter{page}{1}          

   \author{Alexander S. Kutsenko
   \and Valentina I. Abramenko
   \and Andrei A. Plotnikov
   }

   \institute{Crimean Astrophysical Observatory, p/o Nauchny, Crimea, 298409, Russia; {\it alex.s.kutsenko@gmail.com}\\
\vs\no
   {\small Received 20xx month day; accepted 20xx month day}}

\abstract{
Using the data on magnetic field maps and continuum intensity for Solar Cycles 23 and 24, we explored 100 active regions (ARs) that produced M5.0 or stronger flares. We  focus on the presence/absence of the emergence of magnetic flux in these ARs 2-3 days before the strong flare onset. We found that 29 ARs in the sample emerged monotonously amidst quiet-Sun. A major emergence of a new magnetic flux within pre-existing AR yielding the formation of a complex flare-productive configuration was observed in another 24 cases. For 30 ARs, an insignificant (in terms of the total magnetic flux of pre-existing AR) emergence of a new magnetic flux within the pre-existing magnetic configuration was observed; for some of them the emergence resulted in a formation of a configuration with a small $\delta$-sunspot. 11 out of 100 ARs exhibited no signatures of magnetic flux emergence during the entire interval of observation. In 6 cases the emergence was in progress when the AR appeared on the Eastern limb, so that the classification and timing of emergence were not possible. We conclude that the recent flux emergence is not a necessary and/or sufficient condition for strong flaring of an AR. The flux emergence rate of analyzed here flare-productive ARs  was compared with that of flare-quiet ARs analyzed in our previous studies. We revealed that the flare-productive ARs tend to display faster emergence than the flare-quiet ones do.
}

   \authorrunning{A. S. Kutsenko, V. I. Abramenko \& A. A. Plotnikov }            
   \titlerunning{Flux emergence prior to strong flares}  

   \maketitle

%
%
\section{Introduction}           
\label{sect:intro}

Most of the solar flares emanate from active regions (ARs) -- sites where strong magnetic fields penetrate from the convection zone to the solar atmosphere. It is widely accepted that the energy released during solar flares is stored in the form of free magnetic energy caused by non-potentiality of magnetic fields \citep[e.g.][]{Melrose1991, Schrijver2005, Fisher2015}.

Statistical studies reveal that the strongest flares mostly occur in ARs with a complex configuration of the magnetic field \citep[][to mention a few]{Mayfield1985, Sammis2000, Toriumi2017a}. For example, \citet{Guo2014} analyzed 3334 ARs observed from 1983 to 2011 and found that 88\% of X-class flares were produced by ARs with $\beta \gamma \delta$ configuration according to the Mount Wilson classification \citep{Hale1919, Kunzel1960, Kunzel1965}. Recall that in a $\delta$-sunspot two umbrae of opposite magnetic polarities are located within a common penumbra and separated by less than 2 degrees \citep{Kunzel1960}. Note, however, that  \citet{Zirin1987} argued that a presence of any $\delta$-configuration is not sufficient for production of a strong flare and both polarities within the $\delta$-spot must be ``substantial'' for that. 
Flare-productive ARs also frequently display patterns of strong, high-gradient magnetic fields along the highly-sheared polarity inversion line \citep[e.g.][]{Zirin1987, Schrijver2007}. A comprehensive description of morphology and other properties of flare-productive ARs can be found in the recent review by \citet[][]{Toriumi2019} and references therein.

Since ARs with $\delta$-spot are the most plausible candidates to ensure strong flares, a lot of attention has been paid to the formation of such magnetic structures from both observational and theoretical points of view. Thus, \citet{Zirin1987} considered the formation and evolution of 21 flare-productive $\delta$-spot groups observed at the Big Bear Solar Observatory. They concluded that there exists three types of $\delta$-spots formation. First, $\delta$-spots may be formed as a result of simultaneous emergence of one or several magnetic dipoles at one place. Such structures are very compact and exhibit a large umbra. The second type is the emergence of a new dipole in the penumbra of a large pre-existing spot. The size of the newly-emerged dipole must be comparable to that of the pre-existing spot in order to produce strong flares. The third type of $\delta$-spot formation is the collision of the leading part of one dipole with the following part of another dipole during the growth of both dipoles. The $\delta$-spots of this type are less flare-prolific as compared to the aforementioned ones. \citet{Zirin1987} also stated that, once formed, the polarities of a $\delta$-spot never separate. In addition, the orientation of the opposite polarities in $\delta$-spots often disobey Hale's polarity law \citep{Hale1925}. 

\citet{Toriumi2017a} analyzed the morphology and other properties of 29 ARs observed between 2010 and 2016. Each AR produced at least one flare stronger than M5.0 level when it was located within the 45-degrees distance from the central meridian. More than 80\% of ARs exhibited a $\delta$-structure and at least 3 ARs disobeyed Hale's polarity law. The authors proposed four scenarios for the formation of a flare-productive AR:

\begin{enumerate}
\item A complex compact AR with a $\delta$-spot is formed via emergence of a single or several magnetic dipoles. The distinguishing feature of these ARs is a long polarity inversion line that extends through the entire AR. The formation of these ARs might be associated with the emergence of a highly-twisted magnetic flux bundle with kink instability \citep[e.g.][]{Tanaka1991, Linton1996, Knizhnik2018}. The possible anti-Hale orientation of such ARs was explained in a simulation of twisted flux tube emergence by \citet{Fan1999}. This scenario resembles the first type of $\delta$-spot following to \citet{Zirin1987}. In total, 11 of 29 ARs were attributed by \citet{Toriumi2017a} to this category.

\item A complex magnetic structure forms as a result of a new magnetic dipole emergence in the close vicinity of the pre-existing spot. The newly-emerged dipole, labelled ``satellite'' by \citet{Toriumi2017a}, is usually smaller than the pre-existing spot. This scenario refers to the second type of $\delta$-spot formation proposed by \citet{Zirin1987} and 15 out of 29 ARs belonged to this category. \citet{Toriumi2017a} suggested that the minor magnetic flux bundle might be connected to the main spot below the surface or might be trapped by the main tube during its raise through the convection zone. This case, as well as all other proposed scenarios, was successfully modelled by \citet{Toriumi2017b}, see also simulations by \citet{Jouve2018, Cheung2019}.

\item The so-called ``quadrupole'' scenario \citep[type III in][]{Zirin1987}, when two magnetic dipoles emerge close to each other simultaneously and the following spot of the western dipole approaches to the leading spot of the eastern dipole forming the common penumbra.    
An excellent example of such a structure is the well-studied NOAA AR 11158 \citep[e.g.][]{Sun2012, Vemareddy2012}. The dipoles forming the AR might be two emerging parts of the same magnetic flux tube: the rising $\Omega$-tube might split into two segments below the surface \citep{Toriumi2014}, or we observe two rising $\Omega$-segments of a coherent emerging tube \citep{Fang2015, Toriumi2017b, Syntelis2019}.

\item A strong flare occurs between two close ARs \citep[``inter-AR'' event as labelled by][]{Toriumi2017a}. This case is similar to the ``quadrupole'' scenario with an exception that in this case the interacting ARs are supposed to be disconnected below the photosphere. There were only two inter-AR events (NOAA ARs 11944 and 12173) in the sample analyzed by \citet{Toriumi2017a}.
\end{enumerate}

A reasonable inference from the above review seems to be a key role of the magnetic flux emergence some time prior to a strong flare. The free magnetic energy released during a solar flare is stored in the electric currents flowing in the corona and the chromosphere of an AR. The manifestation of these electric currents is non-potentiality of magnetic field in the AR. \citet{Schrijver2005} analyzed the non-potentiality in 95 active regions by comparing extreme ultraviolet images and potential field extrapolations based on photospheric magnetograms. They found approximately one third of analyzed ARs (36 out of 95) to exhibit clear signatures of non-potentiality. These ARs produced stronger and more frequent flares as compared to near-potential ones. \citet{Schrijver2005} found that the majority of non-potential ARs (80-85\%) exhibited intense emergence of a new magnetic flux with the formation of a meandering or fragmented polarity inversion line or the emergence of a new magnetic flux within or in the vicinity of a pre-existing magnetic structure. The emergence was in progress or took place within the last day. On the other hand, the authors found that in 43\% of cases the non-potentiality was observed in ARs with rapidly evolving (touching or cancelling within last 30 hours) opposite-polarity magnetic concentrations of high flux density. Although \citet{Schrijver2005} concluded that non-potentiality in the corona is mostly driven by a complex and large emergence of the magnetic flux, their results suggest that electric currents can be also generated due to shearing/twisting or cancellation of the magnetic flux. Note also that, according to \citet{Schrijver2005}, electric currents injected during the magnetic flux emergence are presumably decay within ~10-30 hours. Therefore, a solar flare is expected to occur within this time interval.

The increase of the frequency of solar flares by a new magnetic flux emerging in the vicinity of a pre-existing one was studied by \citet{Fu2016}. The authors found the enhancement of M- and X-class flaring in pre-existing ARs, which lasts for a day after the emergence of a new magnetic flux in a close vicinity. The same physics -- the interaction between the pre-existing and newly emerged magnetic flux in the corona -- underlie the eruption of a quiescent filaments and initiation of a coronal mass ejection by emerging ARs.

Although the emergence of magnetic flux is studied extensively in conjunction with the appearance of strong flares and eruptions \citep[see, e.g., a review by][]{Schrijver2009}, much less attention is paid to the flux emergence rate of flare-productive ARs. As for observations, \citet{Giovanelli1939} probably was the first one who showed that emerging ARs with higher area growth rate (that is equivalent to higher flux emergence rate) exhibit higher probability of flare occurrence. Numerical simulations predict relatively high values of the flux emergence rate when considering the emergence of a highly-twisted magnetic tube presumably forming $\delta$-spot due to kink instability \citep[e.g.][]{Toriumi2017b, Knizhnik2018}.

In the present study, we aim to test this theoretical deduction regarding faster emergence of flare-prolific ARs. We also intend to reveal whether a recent magnetic flux emergence is necessary for the occurrence of a flare. First, we select ARs that produced M5.0 and larger flares during Solar Cycles 23 and 24. Then, we analyze the magnetic flux emergence in the ARs prior to flares. Finally, we estimate the flux emergence rate and compare it to the flux emergence of flare-quiet ARs.

Note that in our previous paper \citep{Kutsenko2021} we detected ARs emerged on the solar disk between 2010 and 2017. Out of all detected 243 ARs, there were only 34 ARs with noticeable flaring and only one of them produced X-class flare (NOAA AR 11158), so that the comparison between emergence and flaring was possible only for a scanty data sample. We found that the faster an AR emerges the higher the flare index is (the Pearson correlation coefficient reaches 0.74).

\section{Data}
\label{sect:data}

Since our goal is to analyze AR magnetic flux variations associated with solar flares, we need uninterruptible data on AR magnetic fields available at relatively high cadence. The first data source we utilized (1996-2010) was the {\it Michelson Doppler Imager} on board the {\it Solar and Heliospheric Observatory} \citep[SOHO/MDI,][]{Scherrer1995}. SOHO/MDI is a full-disk filtergraph that observes the photospheric Ni~\textsc{i} 6768 \AA\ spectral line at four wavelength positions. The magnetic field is proportional to the difference between the spectral line centres in right- and left-circular polarizations. The derived magnetic field maps are 1024$\times$1024 pixels line-of-sight magnetograms with the pixel size of 2$\times$2 arcsec\textsuperscript{2}. The data are available at 96 minute cadence.

The second source of magnetic field data (2010-2017) was the {\it Helioseismic and Magnetic Imager} \citep{Schou2012} on board the {\it Solar Dynamics Observatory} \citep[SDO/HMI,][]{Pesnell2012}. SDO/HMI observes the Fe~\textsc{i} 6173 \AA\ photospheric spectral line at six wavelength positions to derive solar magnetic field maps. The size of the full-disk maps is 4096$\times$4096 pixels with the pixel size of 0.5$\times$0.5 arcsec$^{2}$. Although the SDO/HMI data of the full magnetic field vector are available, we decided to use SDO/HMI line-of-sight magnetograms of 12 minute cadence due to the following reasons. First, to treat SOHO/MDI and SDO/HMI data by the same procedures to ensure the homogeneity of the derived magnetic flux values. Second, to ensure the consistency of the data reduction techniques applied in this work and in our previous works on AR emergence \citep{Kutsenko2019, Kutsenko2021} in order to perform the comparison of the results. The SDO/HMI algorithm for the line-of-sight magnetic field calculations is similar to that for SOHO/MDI data.

For the analysis we selected ARs that produced M5.0 or stronger flares while the AR was located inside the longitudinal interval (-35, +65) $\degr$. We suppose that the AR magnetic flux can be reliably estimated when the AR is no farther than 60$\degr$ from the central meridian. To ensure 2-3 days of reliable flux measurements before the flare onset, we have to shift the Eastern longitudinal limit to the West by approximately 25$\degr$.

For each AR we compiled the data cube of magnetograms spanning the time interval of the AR's passage across the solar disk. 
For emerging ARs we also tracked the quiet-Sun area where the AR to be appeared. Prior to calculation of the total unsigned magnetic flux, each magnetogram was $\mu$-corrected for the projection \citep{Leka2017}. The magnetic flux density in each magnetogram pixel was divided by the cosine of the angle $\mu$ between the line-of-sight and the vector pointing from the centre of the Sun to the pixel. According to \citet{Leka2017} the $\mu$-correction in each pixel allows us to improve an estimation of the total unsigned magnetic flux from line-of-sight magnetograms. 

The total unsigned magnetic flux was calculated as $\Phi = \sum |B_{i}|\Delta s_i$, where $ B_{i}$ and $\Delta s_i$ are the $\mu$-corrected magnetic flux density and area of $i^{th}$ pixel, respectively. The summation was performed over pixels where the magnetic flux density exceeded the triple noise level $B_{noise}$. The noise level is $B_{noise}=6$ Mx~cm\textsuperscript{-2} for SDO/HMI and $B_{noise}\approx16$ Mx~cm\textsuperscript{-2} for SOHO/MDI magnetograms \citep{Liu2012}. Consequently, the threshold $B_{th}$ was set to 18 Mx~cm\textsuperscript{-2} for SDO/HMI and to 50 Mx~cm\textsuperscript{-2} for SOHO/MDI data. The uncertainty of the calculated unsigned magnetic flux was estimated as $\Phi_{error} = nB_{th}\Delta s$, where $n$ is the number of pixels over which the calculation of magnetic flux was performed.

\begin{figure}
	\includegraphics[width= 0.6\linewidth]{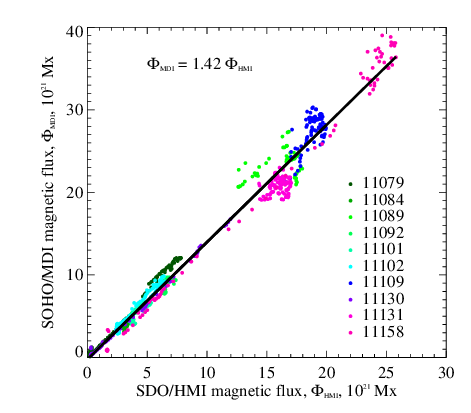}
    \caption{Comparison of the total unsigned magnetic flux calculated using SOHO/MDI and SDO/HMI line-of-sight magnetograms. The magnetic fluxes were calculated from almost co-temporary SOHO/MDI and SDO/HMI magnetograms for ten ARs observed between 2010 June and 2011 February. NOAA numbers of ARs are shown in the plot. Black thick line shows the best linear fit to the data points.}
    \label{figure_method_cc}
\end{figure}

Using SOHO/MDI and SDO/HMI data allowed us to span Solar Cycles 23 and 24 covering the time interval between 1996 and 2017 and to get a large sample of flare-productive ARs. However, the instruments use different spectral lines and have different spatial resolutions. Therefore, in order to get consistent homogeneous estimations of the magnetic flux and flux emergence rate, we performed a cross-calibration between SOHO/MDI and SDO/HMI. 
Fig.~\ref{figure_method_cc} shows SOHO/MDI versus SDO/HMI total magnetic fluxes for randomly selected ten ARs observed between 2010 May and 2011 February. Each point in the plot corresponds to the magnetic flux of a particular AR (coded with different colours) at time points with 96 minute cadence. SOHO/MDI and almost co-temporary SDO/HMI magnetograms of the same region of the solar surface were used to calculate the data points. The linear fitting of the distribution yields the relationship between SOHO/MDI and SDO/HMI magnetic fluxes $\Phi_{MDI} = 1.42~\Phi_{HMI}$. The slope of the fitting line 1.42 is consistent with the value 1.40 by \citet{Liu2012} who compared magnetic flux densities in SOHO/MDI and processed SDO/HMI magnetograms. In the rest of this work, SOHO/MDI magnetic fluxes are divided by a factor of 1.42 to get better consistency with SDO/HMI data.

\begin{figure*}
	\includegraphics[width=\linewidth]{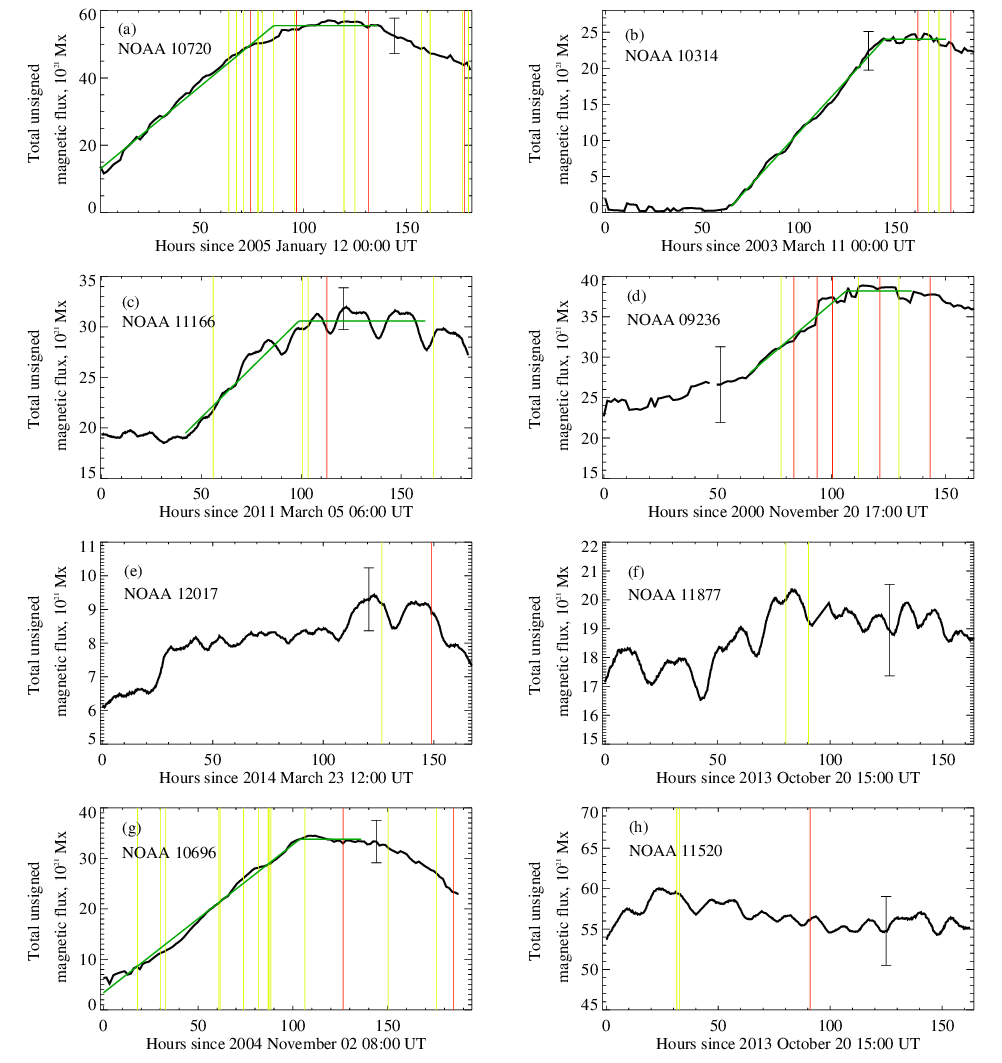}
	\caption{
		Variations of the total unsigned magnetic flux in NOAA ARs 10720 (a), 10314 (b), 11166 (c), 09236 (d), 12017 (e), 11877 (f), 10696 (g), 11520 (h). Vertical lines mark M- (yellow) or X-class (red) flares occurred in the AR. Where applicable, the green line shows the best piece-wise approximation of the emergence part and of the peak pattern by a piece-wise linear fit (see text). Typical errors of the flux measurements are shown in each panel.
	}
	\label{figure_method_fluxes}
\end{figure*}

The total unsigned magnetic flux variations of eight ARs are shown in Fig.~\ref{figure_method_fluxes}.  Diurnal oscillations of the magnetic flux in panels (c),(f), and (h) are caused by well-known artefacts of SDO/HMI instrument \citep[e.g.][]{Liu2012, Kutsenko2016}. We did not take any precautions against this artefact since its influence on the derived values is minor. Error bars in each panel show typical errors of the magnetic flux measurements. 

For ARs with well pronounced emergence stage, the linear fitting of increasing and maximum patterns of the magnetic flux curve by a linear piecewise continuous function \citep[see][]{Kutsenko2019} was performed (green lines in panels (a)--(d) and (g) of Fig.~\ref{figure_method_fluxes}). The fitting function had two segments one of which (left-handed one in Fig.~\ref{figure_method_fluxes}(a-d, g) can have an arbitrary slope. The second segment had to be horizontal. The slope of the left-handed segment yields information on the flux emergence rate, $R_{av}$, in an AR. The error of the flux emergence rate measurements was evaluated as the uncertainty of the slope of the linear fit. The level of the second segment corresponded to the peak magnetic flux, $\Phi_{max}$, of the AR. Note that this fitting was performed only to those magnetic flux curves where (i) a clear emergence was observed in both magnetograms and continuum intensity images, and (ii) the amount of the newly emerged flux exceeded the uncertainty of the magnetic flux estimation.

Vertical lines in Fig.~\ref{figure_method_fluxes} denote the occurrence of M (yellow) and/or X (red) class flares in the AR. The data on soft X-ray flare magnitudes taken by {\it Geostationary Operational Environmental Satellites} (GOES) were retrieved from National Centers For Environmental Information of NOAA\footnote{available at https://www.ngdc.noaa.gov/stp/space-weather/solar-data/}. 

To quantify the flaring productivity of an AR, we calculated a flare index (FI) introduced in \citet{Abramenko2005}:

\begin{equation}
\mathrm{FI} = (100S^{(X)} + 10S^{(M)} + 1.0S^{(C)} + 0.1S^{(B)})/\tau,
\label{eq_fi}
\end{equation}
where $S^{(j)} = \sum_{i=1}^{N_{j}} I_{i}^{j}$ is the sum of \textit{GOES} magnitudes and  $N_{j}$ is the number of flares of a certain class, $\tau$ is the total time interval of AR observation in days. For ARs that passed across the solar disk from the Eastern to the Western limb $\tau$ was set to a half period of solar rotation $\tau = 13.5$ days. For emerging ARs $\tau$ was set to the actual interval of AR presence on the disk.

Beside magnetograms and continuum intensity images, we also utilized UV images acquired by the {\it Extreme-Ultraviolet Imaging Telescope} \citep{Delaboudiniere1995} on board SOHO and by the {\it Atmospheric Imaging Assembly} \citep{Lemen2012} on board SDO to analyze magnetic connections and to locate flares within ARs.

\section{Results}
\label{sect:results}

In total, in this work we analyzed 100 ARs observed between 1996 July and 2017 September which produced strong flares (stronger than M5.0). Actually, we revealed that the total number of ARs that produced M5.0 or stronger flares during this time interval exceeds 150. However, we discarded about one third of them due to either the selection rules described above or gaps in the magnetic field data.   

The list of ARs and their parameters is presented in Table~\ref{table_ar_list} in the Appendix. The common property of most of the ARs was the presence of opposite magnetic polarities located in the close vicinity from each other, although for some of them these polarities were weak to form pores or spots. Only 11 of 100 ARs formally exhibited no $\delta$-spots. In three cases a complex magnetic structure was formed as a result of interaction between two distinct ARs, namely in NOAA ARs 08647, 08674, and 09893 (see the last column in Table~\ref{table_ar_list}).

\subsection{The Relationship Between Flaring and Flux Emergence}
\label{sec_types}

\begin{figure*}
	\includegraphics[width=\linewidth]{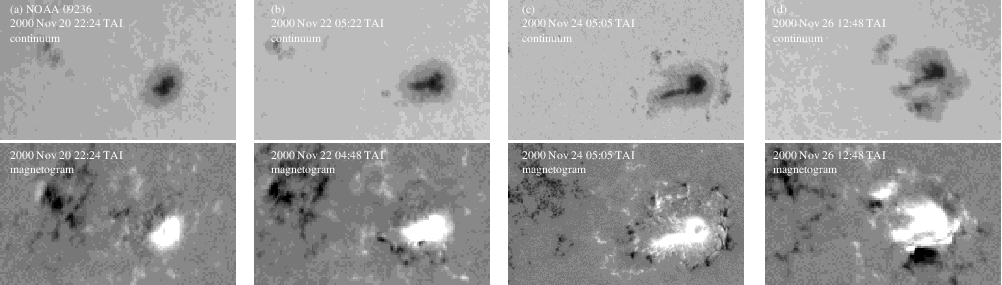}
	\caption
	{
		An example of Type II emergence: Evolution of continuum intensity (upper panels) and of longitudinal magnetic field (lower panels) of NOAA AR 09236. The time stamps are shown in each panel. An intense emergence started in the AR on 2000 November 22 (panels b-d) resulting in the formation of a flaring $\delta$-structure (panel d). The field-of-view of the maps is 125~arcsec$\times$75~arcsec. Magnetograms are scaled from -1000 Mx~cm$^{-2}$ (black) to 1000 Mx~cm$^{-2}$ (white).
	}
	
	\label{figure_09236}
\end{figure*}

\begin{figure*}
	\includegraphics[width=\linewidth]{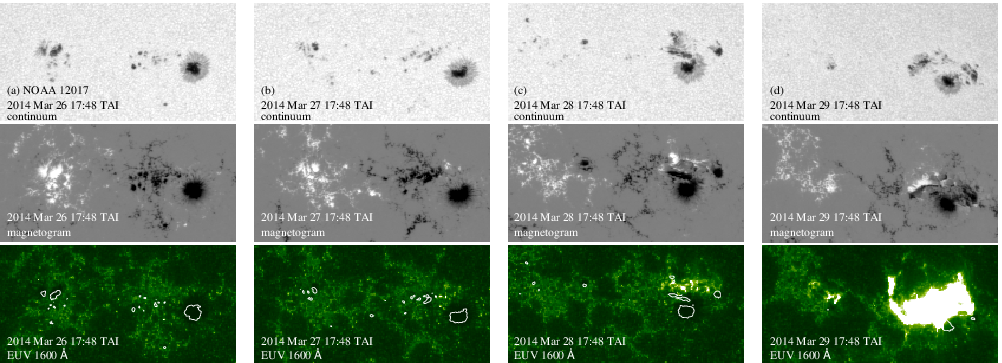}
	\caption{
		An example of Type III emergence: Evolution of continuum intensity (upper panels), longitudinal magnetic field (middle panels), and SDO/AIA 1600 \AA\ intensity (lower panels) of NOAA AR 12017. The time stamps are shown in each panel. White contours in the lower panels show -1000  Mx~cm$^{-2}$ and 1000 Mx~cm$^{-2}$ isolines of the magnetic field. The emergence of a small magnetic dipole in the vicinity of the leading polarity (panels b-c) led to the formation of complex $\delta$-structure (panel d). Lower panel (d) show X1.0 flare on 2014 March 29. The field-of-view of the maps is 100~arcsec$\times$50~arcsec. Magnetograms are scaled from -1000 Mx~cm$^{-2}$ (black) to 1000 Mx~cm$^{-2}$ (white).
	}
	\label{figure_12017}
\end{figure*}

\begin{figure*}
	\includegraphics[width=\linewidth]{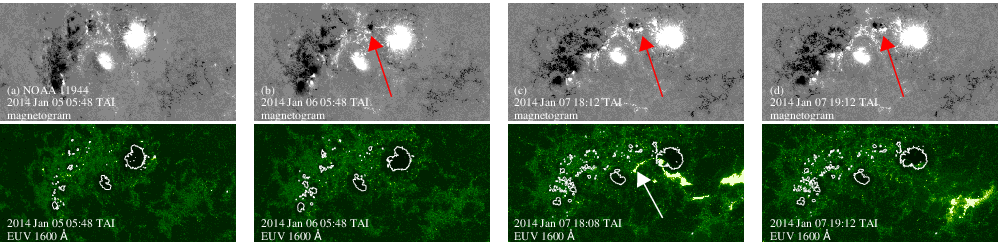}
	\caption{
		An example of Type III emergence: Evolution of longitudinal magnetic field (upper panels) and of SDO/AIA 1600 \AA\ intensity (lower panels) of NOAA AR 11944. The time stamps are shown in each panel. White contours in the lower panels show -1000  Mx~cm$^{-2}$ and 1000 Mx~cm$^{-2}$ isolines of the magnetic field. The emergence of a small magnetic dipole in the vicinity of the leading polarity is shown by red arrows in panels (b)-(d). Lower panels (c) and (d) show X1.2 flare on 2014 January 07. The white arrow in the lower panel (c) points the footpoint of the flare ribbon located near the newly-emerged magnetic dipole. The field-of-view of the maps is 250~arcsec$\times$125~arcsec. Magnetograms are scaled from -1000 Mx~cm$^{-2}$ (black) to 1000 Mx~cm$^{-2}$ (white).
	}
	\label{figure_11944}
\end{figure*}

A thorough analysis of the magnetic flux variations, magnetic field maps, and continuum images of flare-productive AR in our set allowed us to conclude that these ARs can be separated into four subsets. These subsets are not directly related to the AR morphology as in \citet{Zirin1987} or \citet{Toriumi2017a}. ARs in different subsets exhibit different behavior of the magnetic flux emergence prior to or during the strongest flares. We will refer to this behavior as the type of emergence and the assigned type for each AR is shown in column 10 of Table~\ref{table_ar_list}. These types are as follows:
\begin{enumerate}
	\item Type I: A complex magnetic structure forming a flare-productive AR emerges amidst a quiet-Sun area. The total magnetic flux increases monotonously without significant interruptions in the growth. Usually the magnetic structure with $\delta$-spots is formed as a result of emergence of multiple interacting magnetic dipoles. Examples of the total magnetic flux variations of such ARs are shown in panels (a), (b), and (g) of Fig.~\ref{figure_method_fluxes}. A strong flaring (M-class flares) may start during the initial phases of emergence as in Fig.~\ref{figure_method_fluxes}g. However, most commonly M- and X-class flares occur as the AR reaches its peak magnetic flux (Fig.~\ref{figure_method_fluxes}a, b). NOAA ARs 11429 and 11158 can be attributed to this type. The number of ARs assigned to this type of emergence behavior was 29 out of 100.
	
	\item Type II: A complex magnetic structure is formed as a result of emergence of a new magnetic flux within the area of a pre-existing AR. Moreover, the unsigned magnetic flux of the newly injected magnetic structure exceeds the uncertainty of the total magnetic flux measurements, i.e. a considerable amount of a new magnetic flux (as compared to the magnetic flux of the pre-existing structure) appears on the surface. In most cases, the AR exhibits no significant flaring prior to the emergence of the new flux.

	Examples of the magnetic flux evolution of this type of ARs is shown in panels (c) and (d) of Fig.~\ref{figure_method_fluxes}. The magnetic flux of an AR remains nearly constant at a certain non-zero level during the initial phase of observations. Then, the total unsigned flux increases as a result of a new emergence. The flaring activity of the AR increases significantly as well. Probably, the most well-known ARs of this type are NOAA ARs 10930 and 12673. NOAA AR 10930 observed in 2006 December exhibited strong activity near the Eastern limb. It produced two X-class flares and started to decay as a unipolar sunspot surrounded by small magnetic features of both polarities. An emergence of a fast-rotating satellite occurred near the main polarity spot \citep[e.g.][]{Zhang2007}. The formation of a highly-sheared $\delta$-sunspot led to the production of two more X-class flares \citep[e.g.][]{Inoue2012}. NOAA AR 12673 showed up in the Eastern limb in 2017 August as a decaying flare-quiet unipolar sunspot. An intense emergence of a new magnetic flux around the sunspot started on 2017 September 03 and lasted for several days. The AR produced a series of M- and X-class flares including the strongest X9.3-class flare of Solar Cycle 24. 
	
	The evolution of continuum intensity and longitudinal magnetic field of one more AR of this type, namely NOAA AR 09236, is shown in Fig.~\ref{figure_09236}. The variations of the magnetic flux are shown in Fig.~\ref{figure_method_fluxes}d. The AR was a flare-quiet bipolar magnetic structure prior to 2000 November 22 (Fig.~\ref{figure_09236}a). A quite intense emergence began around the leading polarity on this date (Fig.~\ref{figure_09236}b). An interesting feature of the emergence is the formation of a circular symmetrical rim of moving magnetic features around the leading sunspot (Fig.~\ref{figure_09236}c). Three X-class flares occurred in the AR during the emergence, which can be seen in Fig.~\ref{figure_method_fluxes}d. Two more X-class flares were produced by the formed $\delta$-sunspot \citep[Fig.~\ref{figure_09236}d, see also][]{Park2013}. 
	
	For ARs of type II,  in the 6-th column of Table~\ref{table_ar_list} (in addition to the peak magnetic flux, $\Phi_{max}$), we also list  the AR magnetic flux, $\Phi_{min}$, measured prior to the observed emergence onset. There are 24 ARs of type II in our sample. In several cases, when an emerging AR appeared at the Eastern limb, we were unable to determine whether this AR emerges amidst a quiet-Sun area or an additional emergence occurs in the ``old'' AR. Therefore, these ARs were assigned to a mixed type I/II. However, the number of such events is only 6 and we suppose that this uncertainty did not affect the results significantly.
	
	\item Type III: Although a fact of emergence is observed within a pre-existing magnetic structure, the flux injection is negligible as compared to the total magnetic flux of the magnetic structure. This type resembles the ``spot-satellite'' scenario in \citet{Toriumi2017a}. Typical variations of the total magnetic flux of type III ARs are shown in panels (e) and (f) in Fig.~\ref{figure_method_fluxes}. Insignificant (as compared to the uncertainty) injection of a new magnetic flux prior to flares can be seen in the figures.
	
	We assume that the emergence of the satellite may play various roles in the flaring. First, the emergence may result in the formation of a small/moderate $\delta$-sunspot that will produce a strong flare. A large amount of the newly-injected flux is not necessary to produce an X-class flare. For instance, NOAA AR 12017 exhibited one of the lowest in our sample magnitude of the total magnetic flux of about 8$\times$10$^{21}$  (Fig.~\ref{figure_method_fluxes}e). The emergence of 1$\times$10$^{21}$ Mx of additional magnetic flux yielded the appearance of a compact small $\delta$-sunspot that resulted in X1.0 flare on 2014 March 29 at 17:48 UT (Fig.~\ref{figure_12017}).
	
	 Second, we suppose that the emergence of a small magnetic dipole within a pre-existing AR may trigger a flare. And the third possibility is that such an emergence will be irrelevant to the oncoming flare. Unfortunately, modern data and numerical resources are not enough to make an adequate decision on the consequences of a small magnetic flux appearance. As an example, let us consider the case of NOAA AR 11944. Emergence of a new magnetic structure in this AR is shown in Fig.~\ref{figure_11944}. The AR exhibited complex mixed-polarity structure in the following part. Two strong flares (M7.2 and X1.2) were produced by the AR on 2014 January 07. Our focus is the inter-AR X1.2 flare between the strong coherent leading part and the dispersed decayed following part of the neighbor AR 11943 \citep[see also][]{Duan2021}. A small magnetic dipole started to emerge between the leading and following parts of NOAA AR 11944 on 2014 January 06 (shown by red arrow in the upper panels of Fig.~\ref{figure_11944}). The Eastern footpoint of the X1.2-flare ribbon was located 10-20 Mm away from the newly-emerged dipole (white arrow in the lower panel (c) in Fig.~\ref{figure_11944}). The dipole could play some role in the triggering of the flare, however, without additional information from numerical simulations and magnetographic measurements on higher levels it seems impossible to decide.  
	 	 
	  Similar emergence of a new magnetic dipole was observed in the largest and one of the most flare-productive AR of Solar Cycle 24 NOAA 12192. Again, there is not enough information to determine the exact influence of the emerging structures on the triggering of flares. In our opinion, in a certain number of cases these emerging structures do trigger strong flares. In total, 30 out of 100 ARs were identified as type III ARs.
	
	\item Type IV: No clear signatures of emergence were observed during the entire interval of observations (i.e., at least several days prior to the strongest flare occurrence). An example of magnetic flux variations of type IV NOAA AR 11520 is shown Fig.~\ref{figure_method_fluxes}h. The AR exhibited almost constant magnetic flux. 
    The X-class flare occurred three days after that magnetic increase. Moreover, flares may occur during the decaying phase of the AR evolution, see the second (right-hand) X-class flare in Fig.~\ref{figure_method_fluxes}g. Magnetograms of several type IV ARs are shown in Fig.~\ref{figure_type_iv}. Each magnetogram shows the presence of interacting opposite magnetic polarities within the AR. Mutual motions of sunspots resulting in shearing of magnetic field and formation of $\delta$-sunspots were observed in these ARs. There are 11 type IV ARs in our sample.

	 We must conclude that, once formed by any suitable mechanism, an AR with a substantial $\delta$-structure is prone to keep the high flaring potential as long as the $\delta$-structure exists: no additional emergence is required to maintain the high flare activity.

\end{enumerate}

\begin{figure*}
	\includegraphics[width=\linewidth]{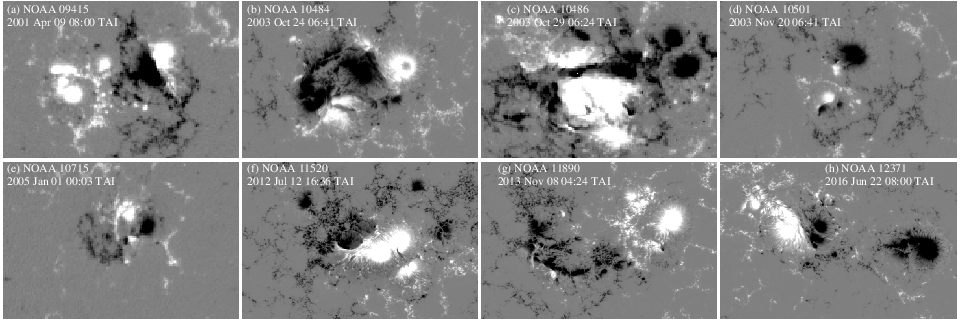}
	\caption{
	Examples of type IV ARs: Longitudinal magnetic field maps of NOAA ARs 09415 (a), 10484 (b), 10486(c), 10501 (d), 10715 (e), 11520 (f), 11890 (g), and 12371 (h). The ARs showed no signatures of magnetic flux emergence during the entire interval of observation and were assigned to type IV. Interacting opposite magnetic polarities are seen in each panel. The field-of-view of the magnetograms is 150~arcsec$\times$100~arcsec. Magnetograms are scaled from -1000 Mx~cm$^{-2}$ (black) to 1000 Mx~cm$^{-2}$ (white).
	}
	\label{figure_type_iv}
\end{figure*}

\subsection{The Flux Emergence Rate of Flare-productive ARs}
\label{sec_fer} 

\begin{figure}
	\includegraphics[width=0.7\columnwidth]{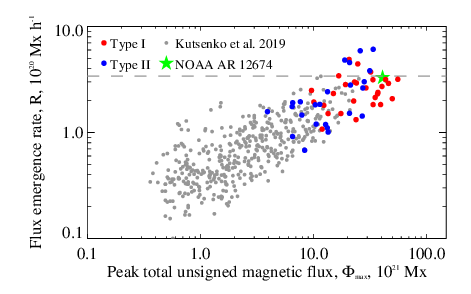}
	\caption{
		The flux emergence rate, $R_{av}$, versus peak magnetic flux, $\Phi_{max}$, distribution for emerging ARs from \citet{Kutsenko2019} (grey circles) and for ARs of type I (red circles) and of type II (blue circles). The green star shows the data point on flare-quiet NOAA AR 12674. Dashed horizontal grey line shows the upper $R_{av}$ limit of the flare-quiet ARs 3.4$\times 10^{20}$ Mx~h$^{-1}$.
	}
	\label{figure_rav_vs_flux}
\end{figure}

As it was mentioned above, numerical simulations suggest highly twisted magnetic flux bundles forming $\delta$-sunspot to exhibit high flux emergence rate during emergence from the convection zone \citep[e.g.][and references therein]{Toriumi2017b, Knizhnik2018}. Most of the ARs in our sample contain $\delta$-sunspots. For type I and type II ARs the data allow us to derive the flux emergence rate $R_{av}$ using the technique presented in \citet{Kutsenko2019}. The relationship between the peak magnetic flux, $\Phi_{max}$, and flux emergence rate for these ARs is shown in Fig.~\ref{figure_rav_vs_flux}. For type II ARs (blue circles), the background flux $\Phi_{min}$ was subtracted before plotting.

Data points from \citet{Kutsenko2019} for 420 emerging dipoles are overplotted with grey circles in Fig.~\ref{figure_rav_vs_flux}. These dipoles are ephemeral and active regions observed between 2010 and 2017, and majority of them were flare-quiet: only 34 of those ARs exhibited flare index exceeding unity. Several ARs, presented in both the samples (in \citet{Kutsenko2019} and in the sample used in this work) were counted once.

In Solar Cycle 24 there was one peculiar AR with very high and stable flux and very low flaring activity. Data for such an AR may allow us to estimate the upper level of the flux emergence rate for flare-quiet AR. This is NOAA AR 12674. EUV images taken by the {\it Extreme Ultraviolet Imager} on board the {\it Solar Terrestrial Relations Observatory} \citep[STEREO,][]{Howard2008, Kaiser2008} allowed us to reveal that this AR started to emerge on the far-side of the Sun on 2017 August 26. By 2017 August 30, the AR was visible near the Eastern limb (~E85)  and exhibited the peak magnetic flux of approximately 4.1$\times 10^{22}$ Mx. We estimated the flux emergence rate to be of about 3.4$\times 10^{20}$ Mx~h$^{-1}$. We overplotted this data point in Fig.~\ref{figure_rav_vs_flux} with a green star. The horizontal dashed line in the figure marks the upper level of the flux emergence rate for all low-flaring ARs, including the largest (in sense of flux) one, AR 12674. Above this level only strong-flaring ARs are observed (however, not all of them).

In general, the diagram in Fig.~\ref{figure_rav_vs_flux} demonstrates that new data (colour circles) are consistent with the published data (grey circles):  the diagram is now extended toward higher magnitudes and the peak total unsigned flux tends to be in a direct proportion with the flux emergence rate. Pearson's correlation coefficient from the combined data set of N=471 events is 0.80$^{+0.03}_{-0.04}$. The slope of the linear regression is the same, 0.48$\pm$0.02. This result indicates that the relationship between $\Phi_{max}$ and  $R_{av}$ found here does not depend on the sample selection.  
 

\begin{figure*}
	\includegraphics[width = \linewidth]{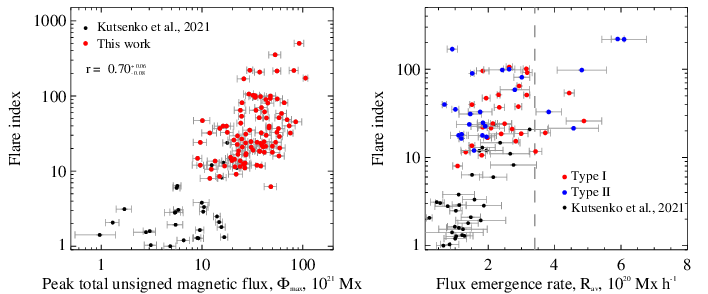}
	\caption{
		Left -- The relationship between flare index, FI, and the peak magnetic flux, $\Phi_{max}$, for ARs analyzed in \citet{Kutsenko2021} (black circles) and in this work (red circles). Measurement uncertainties are shown in the plot. Pearson's correlation coefficient is $r=0.70 ^{+0.06}_{-0.08}$. Right -- The relationship between flare index, FI, and the flux emergence rate, $R_{av}$, for ARs analyzed in \citet{Kutsenko2021} (black circles) and for ARs of type I (red circles) and of type II (blue circles). Measurement uncertainties are shown in the plot. A vertical dashed grey line shows the maximum flux emergence rate $3.4\times 10^{20}$ Mx~h$^{-1}$ measured in flare-quiet ARs.
	}
	\label{figure_fi_vs_all}
\end{figure*}

Fig.~\ref{figure_fi_vs_all} shows the relationship between flare index and peak magnetic flux of all ARs in our sample (left panel) and the relationship between flare index and flux emergence rate of ARs of types I and II (right panel). In both panels we also overplotted data points on ARs (black circles) analyzed in \citet{Kutsenko2021}. These data points are 34 emerging ARs with the flare index exceeding unity.

The left panel in Fig.~\ref{figure_fi_vs_all} confirms a well-known relationship between the peak magnetic flux and flare productivity of ARs: larger ARs are capable of producing stronger flares. Pearson's linear correlation coefficient of the distribution is $r=0.70 ^{+0.06}_{-0.08}$, where the confidence interval is calculated for 95\% confidence level. Although larger ARs tend to produce stronger flares, this relationship is hardly appropriate for reliable flare forecast. Taking into account logarithmic scale on both axes of the plot in the left panel of Fig.~\ref{figure_fi_vs_all}, we may conclude that ARs of similar peak magnetic flux may exhibit flare indices varying in an order of magnitude. High peak magnetic flux is a favourable but not necessary parameter of a flare-productive AR. For instance, NOAA AR 09087 exhibited relatively high peak magnetic flux of about 5.5$\times 10^{22}$ Mx. The strongest flare occurred in this AR was a M6.4 event on 2000 July 19 (see column 7 in Table~\ref{table_ar_list}). Meanwhile, relatively weak NOAA ARs 09511 and 12017 with $\Phi_{max} \approx 9 \times 10^{21}$ Mx produced X1.2 and X1.0 flares on 2001 June 23 and 2014 March 29, respectively.

The relationship between flare index and the flux emergence rate shown in the right panel of Fig.~\ref{figure_fi_vs_all} supports our results reported in \citet{Kutsenko2021}: fast-emerging ARs exhibit, in general, higher flare productivity, the correlation coefficient from the combined data set (N=84) is 0.61$^{+0.12}_{-0.15}$. The similar relationship (with less inclined slope) seems to be valid for the separate subset of type II ARs (blue circles). 

The vertical dashed line shows the $R_{av} = 3.4\times 10^{20}$ Mx~h$^{-1}$ magnitude, the upper level of the flux emergence rate for flare-quiet ARs.  To the right from this segment we observe a well pronounced subset of strong-flaring and fast-emerging ARs with a clear proportionality between FI and $R_{av}$. However, the number of such ARs in our sample is less than 10\%. 

The data points in the right panel of Fig.~\ref{figure_fi_vs_all} are located predominantly above the main diagonal of the plot suggesting that fast-emerging ARs must exhibit high flare-productivity. At the same time, ARs emerging at gradual rates may be flare-quiet or may become flare-productive as well. So, for the entire data set, the relationship between the flux emergence rate and flare index is more complicated than a simple linear regression. Non-linear processes definitely contribute to the formation of flaring capabilities.   

\section{Conclusions and Discussion}
\label{sect:conclusion}

We used SOHO/MDI and SDO/HMI data to analyze 100 ARs that produced M5.0 or stronger flares during Solar Cycles 23 and 24. Our focus was an investigation of the observable magnetic flux emergence events during an interval of approximately 2-3 days before the flaring onset in the AR.  We studied both qualitative and quantitative aspects of the magnetic flux emergence in these flare-productive ARs.

We found that ARs may be sorted out into four types with respect to the emergence of a magnetic flux prior or during strongest flares. Type I consists of the ARs that emerged amidst a quiet-Sun area and started to launch strong flares immediately after (or even before) the magnetic flux reached its peak magnitude.
 The monotonous emergence most often results in the formation of a complex magnetic structure with a $\delta$-spot. Approximately one third (29\%) of all ARs in the sample were assigned to type I.

Emergence of a new magnetic flux within a pre-existing AR was denoted as type II. In most cases, a complex magnetic configuration is formed as a result of this new emergence. The amount of the new magnetic flux is significant as compared to the pre-existing magnetic flux. For instance, the most flare-productive NOAA AR 12673 in Solar Cycle 24 was assigned to type II. Another quarter of all ARs (24\%) were identified as ARs of type II.

For one third of all analyzed ARs (30\%), the emergence of only a small amount of a new magnetic flux was observed. These ARs were labelled as those of type III. Insignificant emergence of a new magnetic flux  may result in the formation of a complex structure tending to produce flares. Interestingly, emergence of a very weak magnetic satellite of about 1$\times 10^{21}$ Mx in the vicinity of main sunspots is enough to form a  small $\delta$-sunspot that could provoke a X-class flare. In our opinion, newly-emerged dipoles may also trigger flares in complex pre-existing ARs. Undoubtedly, emerging dipoles may also play no role in the occurrence of a flare.

A large number of type III ARs in the sample are not promising for the problem of the solar flare forecast. In general, the duration of emergence is proportional to the amount of emerging magnetic flux. Therefore, emergence of a relatively weak satellite in a pre-existing flare-quiet AR may occur within hours. This emergence may result in a complication of the magnetic configuration yielding a start for strong flaring. This means a decrease of the time interval for the reliable prediction down to hours. Probably, new methods for early detection of emerging magnetic flux, for example, by means of helioseismology \citep[e.g.][]{Birch2019, Dhuri2020}, could increase the forecast interval. The distortion of the electric current system of a pre-existing AR by an emerging satellite, which was discussed in \citet{Kutsenko2018}, could also be used in the forecast.

Finally, the type IV ARs are the strong-flaring ARs with no signatures of the new flux emergence during the entire interval of observations. It means that strong flares could occur at least three days after any emergence. All of type IV ARs were characterized by interacting opposite magnetic polarities (see Fig.~\ref{figure_type_iv}). Two of the strongest flares of Solar Cycle 23 belong to this type: NOAA ARs 09415 (X14.4, Fig.~\ref{figure_type_iv}a) and 10486 (X17.0, Fig.~\ref{figure_type_iv}c). Although ARs of type IV are not numerous (only 11 out of 100 ARs), the existence of these ARs implies the following conclusion: the emergence is not a necessary ingredient for an AR to produce a powerful flare; once formed by any scenario, an AR may keep the potential for flaring as long as favorable conditions are met.

To this end, we conclude that
\begin{enumerate}
	\item In 29\% of cases the emergence was observed from the zero background;
	\item In 24\% of cases the major emergence was observed in the pre-existing AR;
	\item In 30\% of cases the minor emergence was observed in the pre-existing AR;
	\item In 11\% of cases no emergence was detected;
	\item For 6\% of cases data did not allow us to judge. 
\end{enumerate}

Our results also support the well-known dependence between the peak magnetic flux and flaring productivity: stronger flares tend to occur in larger ARs. However, this dependence is just a tendency rather than a strong rule. We suppose that the capability of an AR to produce a strong flare is determined to a great extent by its morphology rather than by its size. The examples of relatively weak NOAA ARs 09511 and 12017 discussed in Section~\ref{sect:results} support this suggestion.

In this work we confirmed our previous findings \citep{Kutsenko2021} regarding the flare productivity and flux emergence rate of ARs: ARs emerging at a higher rate tend to produce stronger flares. Although flare-productive ARs exhibit higher flux emergence rate, its consistent difference from $R_{av}$ of flare-quiet ARs is not well pronounced. Flare-productive and flare-quiet ARs rather form a continuous  $R_{av}$ versus $\Phi_{max}$ distribution with flare-productive ARs being mostly located at the higher peak magnetic flux end of the distribution.

The FI versus $R_{av}$ scatter plot (the right panel of Fig.~\ref{figure_fi_vs_all}) suggests that most points lie above the main diagonal, i.e. flux emergence rate defines lower limit of the flare productivity. In other words, ARs emerging at a very high rate must be flare-productive. Gradually emerging ARs may exhibit both high and low flaring productivity. This distribution resembles the relationship between twist and flux emergence rate of ARs found in \citet[][see fig.~3]{Kutsenko2019}. Fast-emerging ARs were found to exhibit strong twist. Gradually-emerging ARs could be either weakly or strongly twisted. Perhaps, both dependencies on the flux emergence rate have the same physical reason: flare-productive ARs with $\delta$-structures are presumably formed as a result of emergence of highly-twisted magnetic flux bundles \citep{Toriumi2017b, Knizhnik2018}.

Very high flux emergence rate can be used as a precursor of strong flare activity of an AR in the future. Our estimates suggest that ARs emerging at a rate higher than $3.4\times 10^{20}$ Mx~h$^{-1}$ will definitely produce strong flares.

However, the number of such ARs is less than 10\% in our sample of flare-productive ARs. The most flare-productive AR of Solar Cycle 24 NOAA 12673 exhibited extremely high flux emergence rate. \citet{Sun2017} estimated the averaged flux emergence rate to be $(4.93^{+0.11}_{-0.13}) \times 10^{20}$~Mx~h$^{-1}$ that is comparable to our value $(5.89 \pm 0.19)\times 10^{20}$ Mx~h$^{-1}$ (see Table~\ref{table_ar_list}). 

We found that a similar flux emergence rate of about  $(6.09 \pm 0.67) \times 10^{20}$~Mx~h$^{-1}$ was observed in NOAA AR 09393. The AR produced one of the strongest flare X20.0 of Solar Cycle 23. Interestingly, both NOAA ARs 09393 and 12673 were classified as type II ARs: in both cases fast emergence of a new magnetic structure was observed in the close vicinity of pre-existing AR.

\begin{acknowledgements}
We are grateful to anonymous referee whose comments helped us to improve the paper. SDO is a mission for NASA’s Living With a Star (LWS) programme. SOHO is a project of international cooperation between ESA and NASA. The SOHO and SDO data were provided by the Joint Science Operation Center (JSOC).
\end{acknowledgements}

\appendix                  
\section{The list of active regions under study and derived parameters}

\onecolumn

\begin{center}
	\begin{longtable}{ccccccccccc}
		\caption
		{
		The list of active regions under study and the derived parameters. NOAA number of AR is listed in column 1. Columns 2 and 3 shows the observational interval in format YYYY.MM.DD. The peak measured total unsigned magnetic flux is listed in column 4. The flux emergence rate for ARs of type I, II, and I/II is shown in column 5. The total magnetic flux measured prior to the emergence in ARs of type II is listed in column 6. The GOES class of the strongest flare observed in the AR within E35-W65 longitudinal interval is listed in column 7. The flare index (FI) of an AR calculated using Eq.~\ref{eq_fi} is shown in column 8. Column 9 shows the most complex observed configuration of an AR according to Mount Wilson classification. Column 10 shows the assigned type of an AR. The last column shows NOAA number of an AR that formed a complex magnetic configuration with the AR under study. 
		} 
		\label{table_ar_list}\\
		
		\hline 
		
		NOAA   & \multicolumn{2}{c}{Obs. dates}    & Peak flux,    & Flux em. rate,        & Min. flux,   & Max. flare & FI & Hale & Type \\
		number &  \multicolumn{2}{c}{(YYYY.MM.DD)} & $\Phi_{max}$, & $R_{av}$,             & $\Phi_{min}$,&            &      & class& \\
			  & Start & End                       & 10$^{21}$ Mx  & 10$^{20}$ Mx~h$^{-1}$ & 10$^{21}$ Mx &            &      &      &  \\
		\hline 
		\endfirsthead
		
		\multicolumn{11}{c}%
		{{\bfseries \tablename\ \thetable{} -- continued from previous page}} \\
		
		\hline
		
		NOAA   & \multicolumn{2}{c}{Obs. dates}    & Peak flux,    & Flux em. rate,        & Min. flux,   & Max. flare & FI & Hale & Type \\
		number &  \multicolumn{2}{c}{(YYYY.MM.DD)} & $\Phi_{max}$, & $R_{av}$,             & $\Phi_{min}$,&            &      & class&  \\
			& Start & End                       & 10$^{21}$ Mx  & 10$^{20}$ Mx~h$^{-1}$ & 10$^{21}$ Mx &            &      &      &  \\
		\hline 
		\endhead
		
		\hline \multicolumn{11}{l}{{Continued on next page}} \\ \hline
		\endfoot
		
		\hline \hline
		\endlastfoot
		
07978&1996.07.04&1996.07.13&$ 16.7\pm  2.6$&$  4.5\pm  0.5$&-- &X2.6& 46.9&$\beta\gamma\delta$&I& \\
08088&1997.09.21&1997.10.01&$ 13.5\pm  2.2$&$  1.5\pm  0.1$&-- &M5.9& 13.6&$\beta\gamma$&I& \\
08100&1997.10.27&1997.11.08&$ 33.7\pm  4.5$&$  4.8\pm  0.7$&$ 14.8\pm  2.6$&X2.1& 97.8&$\beta\gamma\delta$&II& \\
08210&1998.04.25&1998.05.07&$ 27.2\pm  3.9$&$  1.5\pm  0.3$&$ 19.3\pm  3.4$&X1.1& 31.1&$\beta\gamma\delta$&II& \\
08647\footnote{The AR formed a complex magnetic configuration with NOAA AR 08645}&1999.07.25&1999.08.05&$ 49.9\pm  9.3$&$  1.1\pm  0.2$&$ 36.7\pm  7.4$&X1.4& 17.9&$\beta$&II\\
08674\footnote{The AR formed a complex magnetic configuration with NOAA AR 08673}&1999.08.20&1999.09.02&$ 85.2\pm 12.7$&-- &-- &X1.1& 45.3&$\beta\gamma\delta$&III\\
08731&1999.10.11&1999.10.23&$ 37.3\pm  3.9$&$  2.4\pm  0.4$&-- &X1.8& 18.5&$\beta\gamma\delta$&I\\
08778&1999.11.22&1999.12.02&$ 23.9\pm  3.8$&$  1.3\pm  0.1$&-- &M6.0& 11.5&$\beta\gamma$&I\\
08806&1999.12.17&1999.12.30&$ 50.0\pm  5.6$&$  2.1\pm  0.4$&-- &M5.3& 21.5&$\beta\gamma\delta$&I\\
08882&2000.02.21&2000.03.05&$ 38.3\pm  4.2$&-- &-- &M6.5& 11.7&$\beta\gamma\delta$&III\\
08910&2000.03.12&2000.03.24&$ 38.8\pm  5.2$&-- &-- &X1.1& 32.4&$\beta\gamma\delta$&III\\
09026&2000.06.01&2000.06.14&$ 38.9\pm  5.5$&-- &-- &X2.3& 70.1&$\beta\gamma\delta$&III\\
09070&2000.07.02&2000.07.14&$ 24.0\pm  3.4$&-- &-- &M5.7& 13.8&$\beta\gamma\delta$&III\\
09077&2000.07.07&2000.07.20&$ 47.6\pm  6.4$&-- &-- &X5.7& 93.0&$\beta\gamma\delta$&III\\
09087&2000.07.14&2000.07.26&$ 55.5\pm  8.4$&$  3.8\pm  0.4$&$ 23.9\pm  5.7$&M6.4& 32.8&$\beta\gamma\delta$&II\\
09090&2000.07.16&2000.07.28&$ 22.4\pm  3.3$&$  1.9\pm  0.4$&$ 15.8\pm  3.3$&M5.5& 22.7&$\beta\gamma$&II\\
09097&2000.07.18&2000.07.30&$ 26.6\pm  3.2$&-- &-- &M8.0& 11.0&$\beta\gamma$&III\\
09165&2000.09.13&2000.09.22&$ 20.7\pm  2.7$&$  4.9\pm  0.5$&-- &M5.9& 25.9&$\beta\delta$&I\\
09236&2000.11.18&2000.11.29&$ 38.1\pm  5.7$&$  2.4\pm  0.6$&$ 24.8\pm  5.0$&X4.0& 98.2&$\beta\gamma$&II\\
09368&2001.03.02&2001.03.12&$ 22.7\pm  3.4$&$  2.0\pm  0.1$&-- &M5.7& 16.7&$\beta\gamma$&I\\
09393&2001.03.23&2001.04.04&$ 81.6\pm  8.3$&$  6.1\pm  0.7$&$ 47.8\pm  8.2$&X1.7&218.6&$\beta\gamma\delta$&II\\
09415&2001.04.03&2001.04.15&$ 37.4\pm  4.2$&-- &-- &X5.6&208.2&$\beta\gamma\delta$&IV\\
09417&2001.04.03&2001.04.14&$ 12.4\pm  1.8$&$  1.8\pm  0.1$&-- &M8.4& 10.6&$\beta\gamma\delta$&I\\
09433&2001.04.18&2001.05.01&$ 73.4\pm 10.7$&-- &-- &M7.8& 40.0&$\beta\gamma\delta$&III\\
09503&2001.06.13&2001.06.25&$ 48.3\pm  6.9$&-- &-- &M6.2&  6.2&$\beta\gamma\delta$&III\\
09511&2001.06.20&2001.06.30&$  9.6\pm  1.4$&$  2.5\pm  0.2$&-- &X1.2& 24.2&$\beta\gamma\delta$&I\\
09591&2001.08.21&2001.09.04&$ 34.3\pm  4.6$&-- &-- &X5.3& 64.6&$\beta\gamma\delta$&III\\
09601&2001.08.28&2001.09.09&$ 53.8\pm  7.5$&-- &-- &M6.0& 24.2&$\beta\gamma\delta$&III\\
09608&2001.09.05&2001.09.17&$ 57.4\pm  9.1$&-- &-- &M9.5& 36.9&$\beta\gamma\delta$&III\\
09628&2001.09.18&2001.10.01&$ 59.1\pm  7.7$&-- &-- &M7.6& 20.3&$\beta\gamma\delta$&III\\
09632&2001.09.20&2001.10.02&$ 26.1\pm  3.2$&$  2.1\pm  0.7$&$ 19.9\pm  3.6$&X2.6& 23.8&$\beta\gamma\delta$&II\\
09658&2001.10.09&2001.10.21&$  9.6\pm  1.6$&$  1.6\pm  0.4$&$  5.7\pm  1.2$&M5.7& 12.0&$\beta\delta$&II\\
09661&2001.10.10&2001.10.23&$ 43.4\pm  5.3$&-- &-- &X1.6& 30.1&$\beta\gamma\delta$&III\\
09672&2001.10.18&2001.10.30&$ 29.8\pm  4.3$&$  1.0\pm  0.1$&$ 16.2\pm  4.0$&X1.3& 35.1&$\beta\gamma\delta$&II\\
09684&2001.10.28&2001.11.09&$ 26.8\pm  3.8$&-- &-- &X1.0& 12.9&$\beta\gamma\delta$&III\\
09687&2001.11.01&2001.11.13&$ 41.4\pm  7.4$&-- &-- &M9.1& 24.7&$\beta\gamma\delta$&III\\
09704&2001.11.14&2001.11.26&$ 25.1\pm  3.2$&-- &-- &M9.9& 21.0&$\beta\gamma\delta$&III\\
09715&2001.11.24&2001.12.06&$ 39.4\pm  4.6$&$  1.8\pm  0.1$&-- &M6.9& 22.0&$\beta\gamma\delta$&I\\
09727&2001.12.03&2001.12.15&$ 41.5\pm  5.9$&$  1.8\pm  0.4$&$ 31.1\pm  5.5$&M5.6& 17.9&$\beta\gamma\delta$&II\\
09733&2001.12.08&2001.12.20&$ 45.0\pm  6.7$&$  1.1\pm  0.2$&-- &X6.2& 80.5&$\beta\gamma\delta$&I/II\\
09742&2001.12.16&2001.12.28&$ 58.1\pm  7.5$&-- &-- &M7.1& 15.9&$\beta\gamma\delta$&III\\
09773&2002.01.04&2002.01.15&$ 42.9\pm  5.8$&$  4.6\pm  0.8$&$ 21.9\pm  5.1$&M9.5& 21.5&$\beta\gamma\delta$&II\\
09866&2002.03.09&2002.03.21&$ 35.1\pm  4.2$&-- &-- &M5.7& 12.1&$\beta\gamma\delta$&IV\\
09893\footnote{The AR formed a complex magnetic configuration with NOAA AR 09901}&2002.04.03&2002.04.16&$ 37.5\pm  5.6$&$  1.2\pm  0.1$&$ 26.9\pm  5.5$&M8.2& 18.4&$\beta\gamma\delta$&II\\
10017&2002.06.27&2002.07.05&$ 17.5\pm  3.2$&$  1.5\pm  0.1$&-- &X1.5& 39.8&$\beta\gamma\delta$&I\\
10030&2002.07.09&2002.07.22&$ 61.6\pm  7.4$&$  2.8\pm  0.4$&$ 40.3\pm  7.5$&X3.0& 58.6&$\beta\gamma\delta$&II\\
10069&2002.08.11&2002.08.23&$ 65.5\pm  5.9$&$  3.0\pm  0.2$&$ 37.3\pm  6.1$&X1.0& 81.1&$\beta\gamma\delta$&II\\
10139&2002.10.03&2002.10.14&$ 26.1\pm  3.6$&$  4.1\pm  0.6$&-- &M5.9& 12.2&$\beta\gamma\delta$&I/II\\
10226&2002.12.13&2002.12.24&$ 40.5\pm  6.1$&$  2.7\pm  0.1$&-- &M6.8& 21.1&$\beta\gamma\delta$&I\\
10314&2003.03.14&2003.03.21&$ 24.1\pm  3.2$&$  2.9\pm  0.1$&-- &X1.5& 64.6&$\beta\gamma\delta$&I\\
10338&2003.04.18&2003.04.28&$ 16.2\pm  3.0$&$  0.7\pm  0.1$&$  7.8\pm  2.2$&M5.1& 39.8&$\beta\gamma\delta$&II\\
10365&2003.05.19&2003.06.01&$ 29.2\pm  3.3$&$  2.6\pm  0.1$&-- &X1.3&106.1&$\beta\gamma\delta$&I\\
10375&2003.06.01&2003.06.13&$ 47.8\pm  5.9$&$  2.6\pm  0.2$&$ 20.4\pm  3.4$&X1.7&100.4&$\beta\gamma\delta$&II\\
10484&2003.10.18&2003.10.30&$ 58.1\pm  5.2$&-- &-- &X1.2& 51.3&$\beta\gamma\delta$&IV\\
10486&2003.10.22&2003.11.04&$ 92.2\pm 10.1$&-- &-- &X17.2&501.4&$\beta\gamma\delta$&IV\\
10501&2003.11.13&2003.11.25&$ 22.5\pm  3.7$&-- &-- &M9.6& 29.5&$\beta\gamma\delta$&IV\\
10540&2004.01.13&2004.01.25&$ 25.8\pm  4.0$&-- &-- &M6.1& 12.6&$\beta\gamma\delta$&III\\
10564&2004.02.21&2004.03.02&$ 35.2\pm  4.2$&$  2.1\pm  0.1$&-- &X1.1& 24.2&$\beta\gamma\delta$&I\\
10646&2004.07.11&2004.07.16&$ 10.1\pm  1.9$&$  1.9\pm  0.1$&-- &M6.7& 47.0&$\beta$&I\\
10649&2004.07.13&2004.07.25&$ 32.2\pm  5.2$&$  1.7\pm  1.8$&-- &X3.6&102.1&$\beta\gamma\delta$&I/II\\
10652&2004.07.17&2004.07.29&$ 69.9\pm  7.5$&$  3.8\pm  1.0$&-- &M9.1& 48.4&$\beta\gamma\delta$&I/II\\
10656&2004.08.07&2004.08.18&$ 55.9\pm  7.1$&$  3.2\pm  0.1$&-- &X1.0& 91.5&$\beta\gamma\delta$&I\\
10691&2004.10.24&2004.11.04&$ 17.8\pm  2.6$&$  1.8\pm  0.4$&$ 11.3\pm  2.1$&X1.2& 32.9&$\beta\gamma$&II\\
10696&2004.11.02&2004.11.12&$ 33.7\pm  3.7$&$  3.1\pm  0.1$&-- &X2.5&101.0&$\beta\gamma\delta$&I\\
10715&2004.12.29&2005.01.10&$ 12.0\pm  2.0$&-- &-- &X1.7& 32.2&$\beta\gamma\delta$&IV\\
10720&2005.01.11&2005.01.22&$ 55.5\pm  4.9$&-- &-- &X7.1&215.3&$\beta\delta$&III\\
10759&2005.05.09&2005.05.21&$ 21.2\pm  3.1$&-- &-- &M8.0& 11.4&$\beta\gamma$&IV\\
10798&2005.08.15&2005.08.25&$ 16.8\pm  2.3$&$  3.4\pm  0.2$&-- &M5.6& 11.7&$\beta\gamma\delta$&I\\
10808&2005.09.08&2005.09.20&$ 53.3\pm  7.2$&-- &-- &X1.7&353.6&$\beta\gamma\delta$&III\\
10826&2005.11.29&2005.12.09&$ 23.1\pm  3.2$&$  3.0\pm  0.2$&-- &M7.8& 18.6&$\beta\gamma\delta$&I\\
10875&2006.04.24&2006.05.06&$ 21.8\pm  3.5$&-- &-- &M7.9&  9.2&$\beta\gamma\delta$&III\\
10930&2006.12.06&2006.12.18&$ 26.0\pm  3.7$&$  0.9\pm  0.2$&$ 19.5\pm  3.3$&X3.4&169.0&$\beta\gamma\delta$&II\\
11045&2010.02.07&2010.02.15&$ 15.0\pm  2.3$&$  2.3\pm  0.2$&-- &M6.4& 37.0&$\beta\gamma\delta$&I\\
11046&2010.02.08&2010.02.20&$ 12.0\pm  2.4$&$  1.1\pm  0.1$&-- &M8.3&  8.0&$\beta\gamma\delta$&I\\
11158&2011.02.11&2011.02.21&$ 24.7\pm  1.6$&$  4.4\pm  0.2$&-- &X2.2& 53.7&$\beta\gamma\delta$&I\\
11166&2011.03.03&2011.03.16&$ 30.6\pm  2.1$&$  1.8\pm  0.2$&$ 19.2\pm  1.9$&X1.5& 24.7&$\beta\gamma\delta$&II\\
11261&2011.07.27&2011.08.08&$ 24.8\pm  2.0$&$  1.1\pm  0.1$&-- &M9.3& 26.7&$\beta\gamma\delta$&I/II\\
11283&2011.08.31&2011.09.12&$ 24.3\pm  2.0$&-- &-- &X2.1& 43.5&$\beta\gamma\delta$&III\\
11429&2012.03.03&2012.03.16&$ 33.8\pm  2.1$&$  1.8\pm  0.2$&-- &X5.4& 95.8&$\beta\gamma\delta$&I\\
11476&2012.05.06&2012.05.18&$ 46.4\pm  3.2$&-- &-- &M5.7& 39.7&$\beta\gamma\delta$&III\\
11515&2012.06.27&2012.07.09&$ 43.0\pm  3.6$&$  1.5\pm  0.1$&$ 22.2\pm  1.9$&M6.2& 89.7&$\beta\gamma\delta$&II\\
11520&2012.07.07&2012.07.19&$ 58.6\pm  4.2$&-- &-- &X1.4& 29.0&$\beta\gamma\delta$&IV\\
11719&2013.04.05&2013.04.18&$ 15.0\pm  1.3$&-- &-- &M6.5&  8.6&$\beta\gamma\delta$&III\\
11877&2013.10.19&2013.10.31&$ 19.8\pm  1.6$&-- &-- &M9.3& 13.9&$\beta\gamma\delta$&III\\
11884&2013.10.26&2013.11.09&$ 21.1\pm  1.8$&-- &-- &M6.3& 14.4&$\beta\gamma\delta$&IV\\
11890&2013.11.03&2013.11.15&$ 41.8\pm  3.5$&-- &-- &X1.1& 61.4&$\beta\gamma\delta$&IV\\
11936&2013.12.24&2014.01.05&$ 27.2\pm  2.6$&$  1.2\pm  0.1$&$ 14.4\pm  1.8$&M9.9& 16.4&$\beta\gamma\delta$&II\\
11944&2014.01.02&2014.01.15&$ 75.4\pm  4.7$&-- &-- &X1.2& 32.1&$\beta\gamma\delta$&III\\
12017&2014.03.23&2014.04.03&$  9.0\pm  0.9$&-- &-- &X1.0& 14.8&$\beta\gamma\delta$&III\\
12036&2014.04.14&2014.04.23&$ 19.3\pm  1.4$&$  2.8\pm  0.1$&-- &M7.3& 15.3&$\beta\gamma$&I\\
12158&2014.09.05&2014.09.18&$ 27.4\pm  2.0$&$  1.9\pm  0.2$&$ 19.8\pm  1.9$&X1.6& 17.2&$\beta\gamma\delta$&II\\
12192&2014.10.18&2014.10.31&$106.6\pm  5.6$&-- &-- &X2.0&173.0&$\beta\gamma\delta$&III\\
12222&2014.11.27&2014.12.09&$ 30.8\pm  2.1$&-- &-- &M6.1& 20.8&$\beta\gamma$&III\\
12241&2014.12.15&2014.12.26&$ 32.2\pm  2.0$&$  3.7\pm  0.2$&-- &M8.7& 19.1&$\beta\gamma\delta$&I\\
12242&2014.12.15&2014.12.24&$ 43.6\pm  2.8$&$  3.2\pm  0.1$&-- &X1.8& 51.0&$\beta\gamma\delta$&I\\
12297&2015.03.07&2015.03.20&$ 25.0\pm  1.7$&$  0.9\pm  0.1$&-- &X2.2& 81.3&$\beta\gamma\delta$&I/II\\
12371&2015.06.17&2015.06.29&$ 38.2\pm  2.5$&-- &-- &M7.9& 23.4&$\beta\gamma\delta$&IV\\
12403&2015.08.19&2015.08.31&$ 46.2\pm  3.2$&$  2.9\pm  0.1$&-- &M5.6& 37.8&$\beta\gamma\delta$&I\\
12422&2015.09.23&2015.10.04&$ 37.0\pm  2.9$&$  2.3\pm  0.0$&-- &M7.6& 51.3&$\beta\gamma\delta$&I\\
12673&2017.08.30&2017.09.10&$ 29.9\pm  2.0$&$  5.9\pm  0.2$&$  3.9\pm  0.6$&X9.3&220.4&$\beta\gamma\delta$&II\\
	\end{longtable}
\end{center}

\bibliographystyle{raa}
\bibliography{ms2024-0009}

\begin{thebibliography}{50}
\providecommand\natexlab[1]{#1}
\providecommand\JournalTitle[1]{#1}

\bibitem[{Abramenko}(2005)]{Abramenko2005}
{Abramenko}, V.~I. 2005, \apj, 629, 1141

\bibitem[{Birch} {et~al.}(2019)]{Birch2019}
{Birch}, A.~C., {Schunker}, H., {Braun}, D.~C., \& {Gizon}, L. 2019, \aap, 628,
  A37

\bibitem[{Cheung} {et~al.}(2019)]{Cheung2019}
{Cheung}, M.~C.~M., {Rempel}, M., {Chintzoglou}, G., {et~al.} 2019, Nature
  Astronomy, 3, 160

\bibitem[{Delaboudini{\`e}re} {et~al.}(1995)]{Delaboudiniere1995}
{Delaboudini{\`e}re}, J.~P., {Artzner}, G.~E., {Brunaud}, J., {et~al.} 1995,
  \solphys, 162, 291

\bibitem[{Dhuri} {et~al.}(2020)]{Dhuri2020}
{Dhuri}, D.~B., {Hanasoge}, S.~M., {Birch}, A.~C., \& {Schunker}, H. 2020,
  \apj, 903, 27

\bibitem[{Duan} {et~al.}(2021)]{Duan2021}
{Duan}, A., {Jiang}, C., {Zou}, P., {Feng}, X., \& {Cui}, J. 2021, \apj, 906,
  45

\bibitem[{Fan} {et~al.}(1999)]{Fan1999}
{Fan}, Y., {Zweibel}, E.~G., {Linton}, M.~G., \& {Fisher}, G.~H. 1999, \apj,
  521, 460

\bibitem[{Fang} \& {Fan}(2015)]{Fang2015}
{Fang}, F., \& {Fan}, Y. 2015, \apj, 806, 79

\bibitem[{Fisher} {et~al.}(2015)]{Fisher2015}
{Fisher}, G.~H., {Abbett}, W.~P., {Bercik}, D.~J., {et~al.} 2015, Space
  Weather, 13, 369

\bibitem[{Fu} \& {Welsch}(2016)]{Fu2016}
{Fu}, Y., \& {Welsch}, B.~T. 2016, \solphys, 291, 383

\bibitem[{Giovanelli}(1939)]{Giovanelli1939}
{Giovanelli}, R.~G. 1939, \apj, 89, 555

\bibitem[{Guo} {et~al.}(2014)]{Guo2014}
{Guo}, J., {Lin}, J., \& {Deng}, Y. 2014, \mnras, 441, 2208

\bibitem[{Hale} {et~al.}(1919)]{Hale1919}
{Hale}, G.~E., {Ellerman}, F., {Nicholson}, S.~B., \& {Joy}, A.~H. 1919, \apj,
  49, 153

\bibitem[{Hale} \& {Nicholson}(1925)]{Hale1925}
{Hale}, G.~E., \& {Nicholson}, S.~B. 1925, \apj, 62, 270

\bibitem[{Howard} {et~al.}(2008)]{Howard2008}
{Howard}, R.~A., {Moses}, J.~D., {Vourlidas}, A., {et~al.} 2008, \ssr, 136, 67

\bibitem[{Inoue} {et~al.}(2012)]{Inoue2012}
{Inoue}, S., {Shiota}, D., {Yamamoto}, T.~T., {et~al.} 2012, \apj, 760, 17

\bibitem[{Jouve} {et~al.}(2018)]{Jouve2018}
{Jouve}, L., {Brun}, A.~S., \& {Aulanier}, G. 2018, \apj, 857, 83

\bibitem[{Kaiser} {et~al.}(2008)]{Kaiser2008}
{Kaiser}, M.~L., {Kucera}, T.~A., {Davila}, J.~M., {et~al.} 2008, \ssr, 136, 5

\bibitem[{Knizhnik} {et~al.}(2018)]{Knizhnik2018}
{Knizhnik}, K.~J., {Linton}, M.~G., \& {DeVore}, C.~R. 2018, \apj, 864, 89

\bibitem[{K{\"u}nzel}(1960)]{Kunzel1960}
{K{\"u}nzel}, H. 1960, Astronomische Nachrichten, 285, 271

\bibitem[{K{\"u}nzel}(1965)]{Kunzel1965}
{K{\"u}nzel}, H. 1965, Astronomische Nachrichten, 288, 177

\bibitem[{Kutsenko} \& {Abramenko}(2016)]{Kutsenko2016}
{Kutsenko}, A.~S., \& {Abramenko}, V.~I. 2016, \solphys, 291, 1613

\bibitem[{Kutsenko} {et~al.}(2021)]{Kutsenko2021}
{Kutsenko}, A.~S., {Abramenko}, V.~I., \& {Kutsenko}, O.~K. 2021, \mnras, 501,
  6076

\bibitem[{Kutsenko} {et~al.}(2018)]{Kutsenko2018}
{Kutsenko}, A.~S., {Abramenko}, V.~I., {Kuzanyan}, K.~M., {Xu}, H., \& {Zhang},
  H. 2018, \mnras, 480, 3780

\bibitem[{Kutsenko} {et~al.}(2019)]{Kutsenko2019}
{Kutsenko}, A.~S., {Abramenko}, V.~I., \& {Pevtsov}, A.~A. 2019, \mnras, 484,
  4393

\bibitem[{Leka} {et~al.}(2017)]{Leka2017}
{Leka}, K.~D., {Barnes}, G., \& {Wagner}, E.~L. 2017, \solphys, 292, 36

\bibitem[{Lemen} {et~al.}(2012)]{Lemen2012}
{Lemen}, J.~R., {Title}, A.~M., {Akin}, D.~J., {et~al.} 2012, \solphys, 275, 17

\bibitem[{Linton} {et~al.}(1996)]{Linton1996}
{Linton}, M.~G., {Longcope}, D.~W., \& {Fisher}, G.~H. 1996, \apj, 469, 954

\bibitem[{Liu} {et~al.}(2012)]{Liu2012}
{Liu}, Y., {Hoeksema}, J.~T., {Scherrer}, P.~H., {et~al.} 2012, \solphys, 279,
  295

\bibitem[{Mayfield} \& {Lawrence}(1985)]{Mayfield1985}
{Mayfield}, E.~B., \& {Lawrence}, J.~K. 1985, \solphys, 96, 293

\bibitem[{Melrose}(1991)]{Melrose1991}
{Melrose}, D.~B. 1991, \apj, 381, 306

\bibitem[{Park} {et~al.}(2013)]{Park2013}
{Park}, S.-H., {Kusano}, K., {Cho}, K.-S., {et~al.} 2013, \apj, 778, 13

\bibitem[{Pesnell} {et~al.}(2012)]{Pesnell2012}
{Pesnell}, W.~D., {Thompson}, B.~J., \& {Chamberlin}, P.~C. 2012, \solphys,
  275, 3

\bibitem[{Sammis} {et~al.}(2000)]{Sammis2000}
{Sammis}, I., {Tang}, F., \& {Zirin}, H. 2000, \apj, 540, 583

\bibitem[{Scherrer} {et~al.}(1995)]{Scherrer1995}
{Scherrer}, P.~H., {Bogart}, R.~S., {Bush}, R.~I., {et~al.} 1995, \solphys,
  162, 129

\bibitem[{Schou} {et~al.}(2012)]{Schou2012}
{Schou}, J., {Scherrer}, P.~H., {Bush}, R.~I., {et~al.} 2012, \solphys, 275,
  229

\bibitem[{Schrijver}(2007)]{Schrijver2007}
{Schrijver}, C.~J. 2007, \apjl, 655, L117

\bibitem[{Schrijver}(2009)]{Schrijver2009}
{Schrijver}, C.~J. 2009, Advances in Space Research, 43, 739

\bibitem[{Schrijver} {et~al.}(2005)]{Schrijver2005}
{Schrijver}, C.~J., {De Rosa}, M.~L., {Title}, A.~M., \& {Metcalf}, T.~R. 2005,
  \apj, 628, 501

\bibitem[{Sun} {et~al.}(2012)]{Sun2012}
{Sun}, X., {Hoeksema}, J.~T., {Liu}, Y., {et~al.} 2012, \apj, 748, 77

\bibitem[{Sun} \& {Norton}(2017)]{Sun2017}
{Sun}, X., \& {Norton}, A.~A. 2017, Research Notes of the American Astronomical
  Society, 1, 24

\bibitem[{Syntelis} {et~al.}(2019)]{Syntelis2019}
{Syntelis}, P., {Lee}, E.~J., {Fairbairn}, C.~W., {Archontis}, V., \& {Hood},
  A.~W. 2019, \aap, 630, A134

\bibitem[{Tanaka}(1991)]{Tanaka1991}
{Tanaka}, K. 1991, \solphys, 136, 133

\bibitem[{Toriumi} {et~al.}(2014)]{Toriumi2014}
{Toriumi}, S., {Iida}, Y., {Kusano}, K., {Bamba}, Y., \& {Imada}, S. 2014,
  \solphys, 289, 3351

\bibitem[{Toriumi} {et~al.}(2017)]{Toriumi2017a}
{Toriumi}, S., {Schrijver}, C.~J., {Harra}, L.~K., {Hudson}, H., \&
  {Nagashima}, K. 2017, \apj, 834, 56

\bibitem[{Toriumi} \& {Takasao}(2017)]{Toriumi2017b}
{Toriumi}, S., \& {Takasao}, S. 2017, \apj, 850, 39

\bibitem[{Toriumi} \& {Wang}(2019)]{Toriumi2019}
{Toriumi}, S., \& {Wang}, H. 2019, Living Reviews in Solar Physics, 16, 3

\bibitem[{Vemareddy} {et~al.}(2012)]{Vemareddy2012}
{Vemareddy}, P., {Ambastha}, A., \& {Maurya}, R.~A. 2012, \apj, 761, 60

\bibitem[{Zhang} {et~al.}(2007)]{Zhang2007}
{Zhang}, J., {Li}, L., \& {Song}, Q. 2007, \apjl, 662, L35

\bibitem[{Zirin} \& {Liggett}(1987)]{Zirin1987}
{Zirin}, H., \& {Liggett}, M.~A. 1987, \solphys, 113, 267

\end{thebibliography}

\label{lastpage}

\end{document}